	\documentclass[apj]{emulateapj}

\begin{document}
	\title{Imaging Spectropolarimetry with IBIS:
		Evolution of Bright Points in the Quiet Sun}
	\shorttitle{IBIS Spectropolarimetry: BP Evolution}
		
	\author{B.~Viticchi\'{e}\altaffilmark{1},
		D.~Del~Moro\altaffilmark{1},
		F.~Berrilli\altaffilmark{1},
		L.~Bellot~Rubio\altaffilmark{2},
		A.~Tritschler\altaffilmark{3}}
	\shortauthors{Viticchi\'{e} et al.}

	\altaffiltext{1}{Dipartimento di Fisica, Universit\`a degli Studi di Roma
		``Tor Vergata'', Via della Ricerca Scientifica 1, I-00133 Rome, 
		Italy}
	\altaffiltext{2}{Instituto de Astrof\'isica de Andaluc\'ia (CSIC), Apdo.\ de 
      		Correos 3004, 18080 Granada, Spain}
	\altaffiltext{3}{National Solar Observatory/Sacramento Peak, P.O. Box 62, Sunspot, NM
		88349}

	\email{bartolomeo.viticchie@roma2.infn.it}

%---- Abstract
	\begin{abstract}
	We present the results from first spectropolarimetric observations of the solar
	photosphere acquired at the Dunn Solar Telescope
	with the Interferometric Bidimensional Spectrometer. 
	Full Stokes profiles were measured in the \ion{Fe}{1} 
	630.15~nm and \ion{Fe}{1} 630.25~nm lines with high spatial and
	spectral resolutions for $53$ minutes, with a Stokes $V$
	noise of $3\cdot10^{-3}$ the continuum intensity level.
	The dataset allows us to study the evolution of several 
	magnetic features associated with G-band bright points in the quiet Sun.
	Here we focus on the analysis of three distinct
	processes, namely the
	coalescence, fragmentation and cancellation of G-band bright points.
	Our analysis is based on a SIR inversion of the Stokes $I$ and $V$ profiles
	of both \ion{Fe}{1} lines.
	The high spatial resolution of the G-band images combined with
	the inversion results helps to interpret the undergoing physical processes.
	The appearance (dissolution) of high-contrast G-band bright points 
	is found to be related to the local increase (decrease) of the magnetic
	filling factor, without appreciable changes in the field strength.
	The cancellation of opposite-polarity bright points can be
	the signature of either magnetic reconnection or the emergence/submergence
	of magnetic loops.
	\end{abstract}

%---- Keywords
	\maketitle

%---- Paper
	\section{Introduction}
	\label{Intro}
	The improvements in accuracy and spatial resolution of modern
	spectropolarimeters have led to a new concept of quiet Sun
	magnetism. It has been shown, for example, that quiet Sun magnetic
	fields have strengths from zero to $2$~kG and that their evolution 
	is closely related to the granular plasma motions
	\citep[e.g.,][]{LinRim99,DomC03}.

	However, the formation and disappearance of magnetic
	concentrations in the quiet Sun are not yet well understood, mainly
	because of the difficulty of obtaining time sequences of polarization
	measurements over large fields-of-view (FOV).
	The recent upgrade of the Interferometric Bidimensional
	Spectrometer \citep[IBIS;][]{Cav06} to a vector polarimeter
	makes it ideally suited to study these processes with 
	high spatial, spectral and temporal resolution.

	Here we analyze IBIS polarimetric measurements showing the temporal
	evolution of kG fields associated with G-band bright points (BPs) as
	these interact with the photospheric plasma. Our aim is to improve the 
	current knowledge about BP evolution in relation to magnetic fields
	\citep{Ber95,BerTit01,Nis03,SanA04}.
	We focus on poorly known processes like the coalescence, 
	fragmentation and cancellation of
	G-band BPs. This study exploits the physical parameters 
	derived from both the observed profiles
	and the inversion results.
	\begin{figure*}[!ht]
	\centering
	\includegraphics[width=14cm]{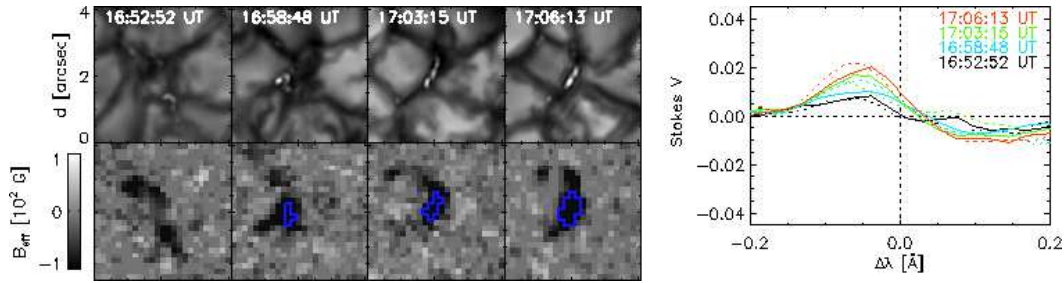}
	\caption{Selection of four instants from the BP coalescence process.
		\textit{First row}: G-band filtergram.
		\textit{Second row}: COG magnetic flux density (images
		are saturated at $100$~G). \textit{Contour plot}:
		kG fields regions as obtained from the inversion analysis
		of Stokes $V$ profiles above $4\cdot\sigma_V$.
		\textit{Right plot}: \ion{Fe}{1} 630.15~nm (\textit{dotted line})
		and \ion{Fe}{1} 630.25~nm (\textit{solid line}) Stokes $V$ profiles
		calculated as an average over a $0.5''\times0.5''$ box around
		the position of the strongest magnetic flux densities for
		each instant. Stokes $V$ profiles are normalized to 
		the continuum intensity.\label{fig1}}
	\end{figure*}
	\begin{figure}[!ht]
	\centering
	\includegraphics[width=7cm]{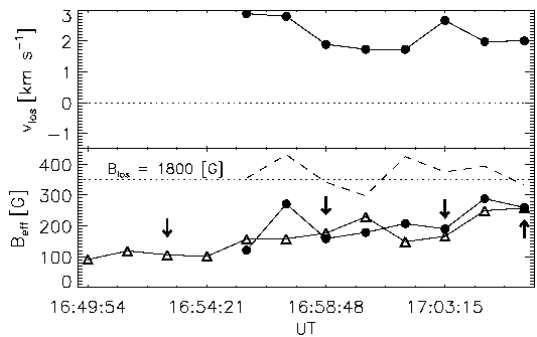}
	\caption{Evolution of relevant quantities for
		the BP coalescence process with full temporal
		resolution. \textit{Upper panel}:
		velocity along the LOS obtained from
		the inversion; positive (negative) values mean 
		photospheric downflows (upflows).
		\textit{Lower panel}: 
		magnetic flux density calculated via COG method (\textit{triangles})
		and from the inversion (\textit{dots});
		we also plot the magnetic field strength obtained from the inversion,
		appropriately rescaled (\textit{dashed line}); the horizontal dotted
		line is a reference level for the field strength.
		The black arrows mark the instants shown in Figure~\ref{fig1}.
		\label{fig1p}}
	\end{figure}

	\section{Observations}
	\label{Data}
	The observations were taken with IBIS at the NSO/Dunn Solar Telescope on
	November 21, 2006 from 16:24 UT to 17:17 UT. The dataset 
	consists of $36$ scans of the \ion{Fe}{1} 630.15~nm and
	\ion{Fe}{1} 630.25~nm lines with $89$ seconds cadence, obtained at 
	disk centre. The two spectral lines are sampled with a total of
	$45$ wavelengths (FWHM~$=2$~pm) on an equidistant grid of $2.3$~pm skipping
	the telluric line in between the two \ion{Fe}{1} lines, but
	sampling the O$_2$ 630.28~nm which allows to set the absolute wavelength
	scale.

	In spectropolarimetric mode, the incoming light to IBIS
	is modulated by a pair of nematic liquid crystal variable retarders 
	placed in a collimated beam in front of
	the field stop of the instrument.
	The light is analyzed by a beam splitter in front of the
	detector, imaging two orthogonal states onto the same chip
	thus allowing for dual-beam spectropolarimetry.
	The modulation is in such a way that at each wavelength position
	six modulation states $I+S$ (and its orthogonal states $I-S$)
	are detected with the following temporal scheme: $S=[+V,-V,+Q,-Q,+U,-U]$.

	The pixel scale of the spectropolarimetric 
	images is $0.18''$, while the integration time per modulation state and
	wavelength was $80$~ms.  For each narrow-band filtergram,
	a simultaneous broad-band ($633.32\pm5$~nm) counterpart was acquired, imaging
	the same FOV with the same exposure time.
	Furthermore, G-band filtergrams ($430.5\pm0.5$~nm) with
	approximately the same FOV, but smaller pixel scale
	($0.037''$) were taken with an exposure time of
	$15$~ms. The seeing during the acquisition run was excellent and
	stable, allowing the adaptive optics system 
	\citep{Rim04} to achieve near diffraction-limited performance.

	The broad-band and G-band images have been restored via Multi-Frame
	Blind Deconvolution \citep[MFBD; ][]{Lof02} to further reduce the
	seeing degradation and obtain a homogeneous resolution in the whole
	$50''\times50''$ FOV.

	The global and local shifts necessary to align and destretch the broad-band images
	with respect to the MFBD restored broadband images have been computed and
	applied onto the spectropolarimetric images. This process reduces the
	seeing-induced crosstalk and makes the spatial resolution of the
	whole spectropolarimetric scan comparable with that of the individual
	narrow-band filtergrams.

	The dataset was then corrected for the systematic wavelength shift
	across the FOV \citep{ReaCav08} caused by the collimated mounting 
	of the Fabry-Perots and for instrumental polarization introduced
	by the telescope and the polarimeter itself.
	The average noise level for Stokes $V$ has been measured to be 
	$\sigma_V=3\cdot10^{-3}$ in units of the continuum intensity.

	The G-band observations are aligned with the broad-band
	images by using grid line targets.
	We then registered the different scans via a correlation
	procedure to eliminate residual global shifts due to
	tracking inaccuracies.
	\begin{figure*}[!ht]
	\centering
	\includegraphics[width=14cm]{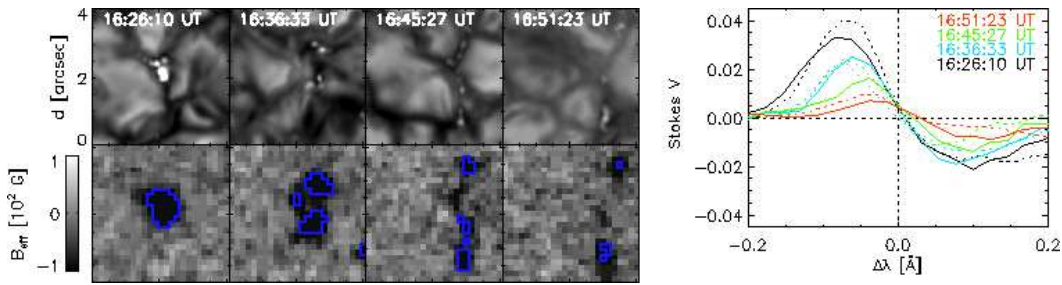}
	\caption{Selection of four instants from the BP fragmentation process.
		Same as Figure~\ref{fig1}.\label{fig2}}
	\end{figure*}
	\begin{figure}[!ht]
	\centering
	\includegraphics[width=7cm]{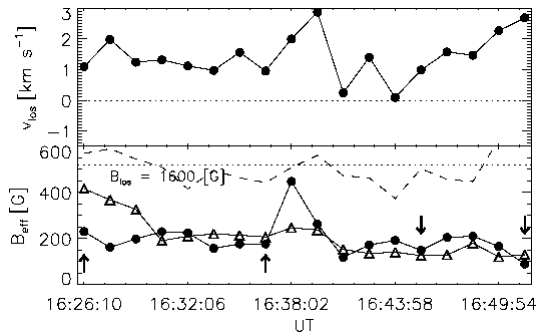}
	\caption{Evolution of relevant quantities for
		the BP fragmentation process with full temporal
		resolution. Same as Figure~\ref{fig1p}.\label{fig2p}}
	\end{figure}
	
	\section{Data analysis}
	\label{Inversion}
	We estimate the longitudinal magnetic flux density applying the centre-of-gravity
	method \citep[COG,][]{ReeSem79} to the observed Stokes $V$ profiles of the 
        \ion{Fe}{1} 630.15~nm line. Unlike the classical magnetograph formula, the COG
	technique does not suffer from saturation effects in the kG regime.

	To determine the intrinsic field strengths and magnetic filling
	factors associated with the G-band BPs we perform an
	inversion of the Stokes $I$ and $V$ profiles of the two lines with the
	SIR code \citep{RuiCTorI92}. We defer the interpretation of the linear
	polarization to future work with a more specific analysis as
	recommended by \citet{Lit08}.

	The inversion is based on two atmospheric components, one magnetized
	and the other field-free. The fractional area of the pixel occupied by
	the former (the magnetic filling factor $f$) is a free parameter of the
	inversion. The temperature stratification of each component is
	modified with two nodes, using the stratification of the Harvard
	Smithsonian Reference Atmosphere \citep{Gin71} as initial guess. The
	line-of-sight (LOS) velocities in the two components as well as the field strength in
	the magnetized component are assumed to be constant with height. The
	stray-light contamination is modeled by averaging the Stokes $I$
	spectra in a region of $1''$ around the inverted profile.
	This average profile is added to the observed spectra
	weighted by a factor $\alpha$; the stray-light factor distribution
	is in good agreement with \cite{OroS07} with typical
	value of $\alpha=80\%$. The macroturbulence and microturbulence
	velocities are set to zero. In the inversion process, the finite
	spectral resolution of the instrument is taken into account using the
	spectral point spread function of IBIS \citep{ReaCav08}.

	Since our goal is to study the evolution of strong kG flux
	concentrations associated with G-band BPs, we invert only those
	profiles whose Stokes $V$ signals are above $4\cdot\sigma_V$ in both \ion{Fe}{1}
	lines; such regions always enclose G-band BPs.
	A total of $53000$ profiles meet this condition, corresponding 
	to approximately $3\%$ of the whole FOV at any step of the time sequence.

	\section{Evolution of Magnetic Features}
	\label{Evo}
	This sequence of high-spectral and
	-spatial resolution observations allows us to study
	the coalescence, fragmentation and cancellation
	of small-scale magnetic structure.
	Examples of these processes are identified
	by inspecting the MFBD-restored G-band time series
	taken as context data in addition to the spectropolarimetric
	observations.

	Figures~\ref{fig1},\ref{fig2} and \ref{fig3} exemplify
	each individual process with a selection of four
	G-band subfields ($4''\times4''$) (first row),
	simultaneous and co-spatial maps of COG magnetic flux density
	(second row) and Stokes $V$ profiles
	(right plot) averaged over a $0.5''\times0.5''$
	box centered on the strongest COG signal.
	The contour lines overlaid on the magnetic flux density
	delimit kG field regions as indicated by the
	inversion of both \ion{Fe}{1} lines.

	Figures~\ref{fig1p} and \ref{fig2p} display
	plots of several quantities relevant to the
	processes shown in Figures~\ref{fig1} and \ref{fig2}
	with full temporal resolution. We report the evolution 
	of the LOS velocity for the
	magnetized component ($v_{los}$), the magnetic flux density 
	($B_{eff}$) and LOS field strength ($B_{los}$).
	$v_{los}$ and $B_{los}$ are outputs of the inversion
	analysis, while $B_{eff}$ has been computed
	from both the inversion as $f\cdot B_{los}$ 
	(dots) and from the COG magnetic flux density
	(triangles).	
	The plotted quantities are averaged in the same
	$0.5''\times0.5''$ box defined above.

	\subsection{Coalescence}
	\label{BPcoal}
	In Figure~\ref{fig1} a BP coalescence process is shown.
	In the first frame a diffuse magnetic signal can be recognized
	in the magnetic flux density images. 
	In the G-band filtergram, scattered, low-contrast bright features associated with 
	the diffuse magnetic signals are recognizable. 
	The photospheric advection field gathers the initially diffuse 
	magnetic features in a small region of about $0.5''$; simultaneously,
	in the same region, a BP emerges in the intergranular lane 
	and kG fields are revealed by the inversion procedure
	(Stokes $V$ profiles emerge above $4\cdot\sigma_V$). 
	
	In a time interval of about $10$~min, the
	magnetic flux density more than doubles  
	(Figure~\ref{fig1p}). 
	Simultaneously, the BP extends over $0.5''$ 
	in the intergranular lane.
	The LOS velocity
	is between $2-3$~km~s$^{-1}$.
	The magnetic field strength
	remains nearly constant at about $1.8$~kG, while the magnetic flux density
	increases in good agreement with the same quantity
	calculated via COG method (Figure~\ref{fig1p}). 
	These results indicate that the
	increase of Stokes $V$ amplitude is due to the increase of
	the magnetic filling factor and not to an
	increase of the field strength; the inversion
	analysis retrieves an increase of $f$ from about $5\%$
	to about $14\%$. A confirmation
	of this result comes from Stokes $V$ amplitudes
	in Figure~\ref{fig1}: 
	since the beginning of the sequence, the Stokes $V$ 
	profiles of the two \ion{Fe}{1} lines have comparable amplitudes,
	so the field is always in the kG regime.
		
	\subsection{Fragmentation}
	\label{BPfrag}
	Figure~\ref{fig2} shows the fragmentation of a BP. In a
	time interval of about $25$ minutes, a high-contrast BP is
	fragmented in many small-scale bright features that move all
	over an intergranular lane with a length of
	about $4''$. The Stokes $V$ signals are reduced in this
	process, but the comparable Stokes $V$ amplitudes of the two
	\ion{Fe}{1} lines indicate again a kG regime.

	Figure~\ref{fig2p} shows that the magnetic field strength
	remains constant at about $1.6$~kG, while the smearing of the
	``progenitor'' BP causes a decrease of the magnetic flux
	density from about $400$~G to about $80$~G. These results
	suggest that the magnetic filling factor is reduced by the
	action of the photospheric advection field; in fact the inversion
	analysis retrieves a decrease of $f$ from about $13\%$
	to about $4\%$.
	The LOS velocity exhibits smaller values than in the BP
	coalescence case; more specifically, it varies around
	$1$~km~s$^{-1}$.

	\subsection{Cancellation}
	\label{BPcanc}
\begin{figure*}[!ht]
	\centering
	\includegraphics[width=14cm]{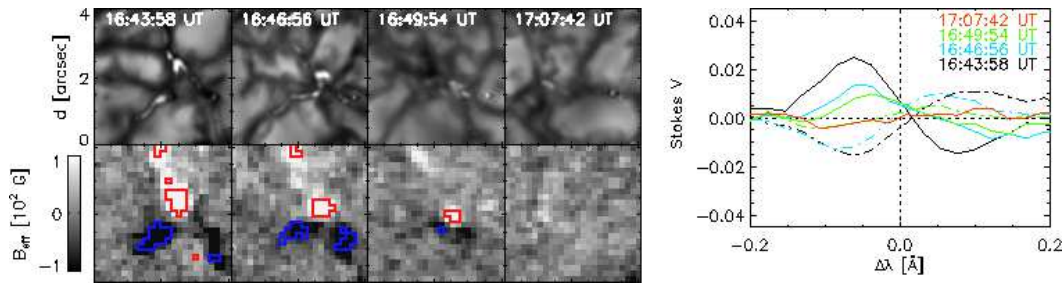}
	\caption{Selection of four instants from the BP cancellation process.
		\textit{First row}: G-band filtergram.
		\textit{Second row}: COG magnetic flux density (images
		are saturated at $100$~G). \textit{Contour plot}:
		kG fields regions as obtained from the inversion analysis
		of Stokes $V$ profiles; positive (\textit{red contours}) and negative 
		(\textit{blue contours}) polarity regions are represented.
		\textit{Right plot}: \ion{Fe}{1} 630.25~nm Stokes $V$ profiles
		calculated as average over a $0.5''\times0.5''$ box around
		the position of the minimum (\textit{solid line}) and 
		maximum (\textit{dot-dashed line}) magnetic flux densities
		for each time instant. Stokes $V$ 
		profiles are normalized to the continuum intensities.
		For 17:07:42 UT a single $0.5''\times0.5''$ box fixed
		at the interaction point is used to calculate the average
		profiles.\label{fig3}}
	\end{figure*}

	The process we present in Figure~\ref{fig3}
	shows the cancellation of opposite polarity BPs converging
	over the same photospheric region.

	The photospheric advection field
	gathers together opposite polarity flux concentrations.
	The result is a rapid disappearance of the G-band BPs 
	and the complete cancellation of the magnetic signals. 
	The whole process lasts about $25$~min.
	In the plot of Figure~\ref{fig3} we report \ion{Fe}{1} 630.25~nm
	Stokes $V$ profiles associated with both polarity BPs.
	From the gradual reduction of the profile amplitudes
	we conclude that a continuous process is
	causing the disappearance of the circular
	polarization signals.

	\section{Discussion and Conclusions}
	\label{Conc}
	Recently, \citet{Nag08} and \citet{BelGon09} presented
	analyses of events
	in which the amplification of Stokes $V$ signal was found to
	be simultaneous to the appearance of bright features in
	continuum and broad band images, respectively.
	\citet{Nag08} performed an inversion of both \ion{Fe}{1} 630~nm
	lines and measured an increase of the magnetic field
	strength associated with very strong redshifts (up to $6$~km~s$^{-1}$).
	The authors showed how Stokes $V$ profiles evolved during
	the amplification process, changing from a sub-kG regime
	(\ion{Fe}{1} 630.15~nm Stokes $V$ amplitude lower than the
	\ion{Fe}{1} 630.25~nm one) to a kG regime (the Stokes $V$ profiles
	of the two lines have comparable amplitudes). 
	All these aspects are compatible with a convective collapse process.
	\citet{BelGon09} also observed an
	increase of the Stokes $V$ signal in the \ion{Fe}{1} 617.3~nm
	line and tentatively suggested the occurrence of a convective
	collapse.

	The approach adopted in the present work shows that the inversion
	of the observed profiles can help to interpret the physical
	process behind such observations.
	In particular, the increase of the filling factor reported in \S~\ref{BPcoal} is 
	induced by the action of the photospheric advection field.
	By the same token, the fragmentation event described in \S~\ref{BPfrag} and associated 
	with a decrease of the kG field filling factor, is produced by
	an opposite driving action of the advection field.
	Furthermore, the more stable velocity in Figure~\ref{fig1p}
	suggests that the gathering of bright features 
	in the coalescence process takes place in a steady downflow,
	while the variability of the velocity in Figure~\ref{fig2p} 
	suggests that such a stability is lost during the fragmentation process.
	In both cases, the contrast of G-band bright points is strictly
	correlated to the fraction of atmosphere filled by kG fields,
	and therefore controlled by the advection field.

	Different scenarios can explain the cancellation process.
	The first is the interaction, via successive magnetic reconnections, 
	between opposite polarity field lines belonging
	to different magnetic bundles.
        A second possible scenario is the emergence of a 
	U-shaped magnetic loop or the submergence of a $\Omega$-shaped loop 
	whose footpoints are observed as BPs.
	In both scenarios, a certain amount of linear
	polarization is expected to be found between the circularly
	polarized regions, during the cancellation phase.
	Contrary to this expectation, IBIS was not able to detect any trace of linear
	polarization. Higher polarimetric accuracy
	seems to be required to investigate this issue.

	We have presented results obtained from first
	observations with IBIS in
	spectropolarimetric mode. They provide new information on
	three distinct and poorly known processes, namely the
	coalescence, fragmentation, and cancellation of flux
	concentrations in the photosphere of the quiet Sun.
 	Our analysis highlights the importance of interpreting Stokes
	profiles via inversion techniques to identify the physical
	mechanisms behind the processes under examination. In this way
	we have been able to associate the appearance (dissolution) of
	high-contrast G-band BPs with a local increase (decrease) of
	the filling factor of photospheric kG fields induced by
	the photospheric advection field. Apparently, the
	field strength does not undergo significant variations in
	those processes (Figures~\ref{fig1p} and \ref{fig2p}). Also,
	we have analyzed the cancellation of opposite-polarity G-band
	BPs. Our results add further data to the recent 
	observation of interactions between granular flows and magnetic elements 
	acquired with \textit{HINODE} \citep[e.g.,][]{Zha09}.
	
      \acknowledgements
	We are very grateful to the anonymous referee for the constructive
	comments and remarks on the manuscript.
	This work was partially supported by the MAE Spettro-Polarimetria Solare 
	Bidimensionale research project, by the
	Agenzia Spaziale Italiana through grant ASI-ESS, by the
	Istituto Nazionale di Astrofisica through grant PRIN-INAF 2007,
	by the Spanish MICINN through project ESP2006-13030-C06-02 and by
	Junta de Andaluc\'{\i}a through project P07-TEP-2687.
	The authors are grateful to the DST observers D. Gilliam, M. Bradford and
	J. Elrod. IBIS was built by INAF-Osservatorio Astrofisico 
	di Arcetri with contributions from the Universit\`a
	di Firenze and the Universit\`a di Roma ``Tor Vergata''. 
	The authors acknowledge F. Cavallini, K. Reardon, 
	and the IBIS team for their invaluable and unselfish support.
	NSO is operated by the Association of Universities for Research 
	in Astronomy, Inc. (AURA), under cooperative agreement with the 
	National Science Foundation.

%---- Bibliography

	\end{document}